\documentclass[10pt,twocolumn]{article}
\usepackage{latexsym}
\usepackage[dvips]{graphicx}
\usepackage{fancyhdr}

\begin{document}

\title{Classical trajectories and quantum supersymmetry}
\author{Piotr Korcyl \\ \small{\emph{M. Smoluchowski Institute of Physics, Jagellonian
University}} \\ \small{\emph{Reymonta 4, 30-059 Krak\'{o}w,
Poland}}}
\date{}

\twocolumn[
\begin{@twocolumnfalse}
      \maketitle
      \begin{abstract}
        We analyze a supersymmetric system with four flat directions. We
        observe several interesting properties, such as the coexistence of
        the discrete and continuous spectrum in the same range of energies.
        We also solve numerically the classical counterpart of this system.
        A similar analysis is then done for an alike, but non-supersymmetric
        system. The comparison of theses classical and quantum results may
        serve as a suggestion about classical manifestations of
        supersymmetry.\\\\

      \end{abstract}
    \end{@twocolumnfalse}

    \vspace{10cm}

    \begin{flushleft}
    TPJU-14/2006 \\ October 2006
    \end{flushleft}
]

\pagebreak

    Great interest has been recently put on the Supersymmetric Yang-Mills
Quantum Mechanics (SYMQM), which results from a dimensional
reduction of the supersymmetric Yang-Mills field theories to a
single point in space. The use of simple numerical methods gives a
good understanding of the $N = 2$, $D = 2$ and $D = 4$ models
\cite{wosiek, w2, w3}. The ultimate goal of such analysis is the $D
= 10$, $SU(N \rightarrow \infty)$ model, which is conjectured to be
in relation with the M-theory \cite{susskind}.

    The system addressed here was used by de Wit, L\"uscher and Nicolai in
\cite{nicolai, nicolai2} as a simplified model of the supermembrane
hamiltonian in an extensive discussion on stability of
supermembranes, structures present in the M-theory. It is one of the
simplest supersymmetric models with flat directions. It thus
contains several interesting properties of SYMQM, such as the
coincidence of the discrete and continuous spectrum. The aim of this
paper is to analyze in detail this system with the method proposed
by Wosiek \cite{wosiek} and to compare it with an analogous, but
non-supersymmetric model. The latter was also already described in
the literature \cite{mertens}, because of its interesting feature:
all its eigenstates are localized, even though the potential is zero
on an unbounded set.

The paper is composed as follows. We start by introducing the
quantum systems: the non-supersymmetric one and the supersymmetric
one, and we continue by presenting their classical counterparts. The
comparison in the quantum regime is based on the analysis of
numerical spectra, therefore we proceed by describing the method
used for calculating spectra, and then give a detailed study of all
symmetries present in theses systems in order to fully understand
the degeneracies which may appear. Consequently we discuss the
results of this comparison. Finally, we turn to the classical regime
and analyze the classical bosonic and supersymmetric trajectories.
Comparing them enables us to rediscover the differences between
quantum systems on the classical level.

\subsection*{Description of the systems} \label{sec:omowienie ukladow}
    In this chapter we introduce the hamiltonians of the considered
models; first the quantum, then the classical one. In the following,
the non-supersymmetric system will be called bosonic, because it
contains only bosonic degrees of freedom.

    \subsubsection*{Quantum systems}
    The quantum bosonic hamiltonian has the form
            \begin{equation}
            \label{eq:h_boz_qm}
            \hat{H}_{bosonic} = \hat{p}^{2}_{x} + \hat{p}^{2}_{y} +
            \hat{x}^{2}\hat{y}^{2}.
            \end{equation}
    The potential $x^2y^2$ has four flat directions. Usually, we expect
that systems with potentials equal to zero on an unbounded set, have
continuous spectrum. Nevertheless, in our case, the spectrum turns
out to be exclusively discrete due to the quantum fluctuations in
the transverse directions. Let's consider a particle moving in one
of the valleys, say $x > 0$. The transverse potential is the
potential of a harmonic oscillator with an equilibrium position at
$y=0$ and frequency proportional to the distance from the center $
\omega \sim |x|$.  The zero-mode energy of such fluctuations
increases linearly with $|x|$, when the particle is moving deep into
the valley. Therefore, the particle is exposed to an effective
potential barrier, which prevents it from escaping. We will now
recall \cite{mertens} the solution of the stationary Schr\"odinger
equation in the Born-Oppenheimer approximation in order to
explicitly show the above. The energy of a quantum harmonic
oscillator of frequency $|x|$ is equal
\[E_x = |x|(n_y + \frac{1}{2}).\]
The wavefunction can be written as a product of two functions
\[ \Psi(x,y) = \alpha (x) \beta_{x,n_y} (y), \]
where $\alpha (x)$ accounts for the onward motion, and
$\beta_{x,n_y}(y)$ describes the transverse harmonic fluctuations
\[ (-\frac{\partial^2}{\partial y^2} + x^2 y^2)\beta_{x,n_y}(y) = 2|x|(n_y+\frac{1}{2}) \beta_{x,n_y}(y). \]
Thus, $\beta_{x,n_y}(y)$ is a well-know eigenfunction of quantum
harmonic oscillator
\[ \beta_{x,n_y}(y) = (\sqrt{\pi} 2^{n_y} n_y ! )^{-1/2} H_{n_y}(\sqrt{2|x|}y) \ e^{-y^2 |x|}.\]
$\alpha(x)$ satisfies a stationary Schr\"odinger equation
\[ (-\frac{\partial^2}{\partial x^2} + 2|x|(n_y + \frac{1}{2})) \alpha (x) = E_{n_x,n_y} \alpha (x). \]
We now proceed independently for $x>0$ i $x<0$. Let us introduce a
new variable $z$
\[ z = (n_y + \frac{1}{2})^{1/3}\Big(x - \frac{E}{2n_y + 1}\Big). \]
The equation for $\alpha(x)$ can be now rewritten in this variable
as
\[ \ddot{\alpha}(z) - 2z \alpha(z) = 0. \]
This is the Airy equation. Its solutions are even and odd Airy's
functions. In order to obtain a full solution we have to match
together solutions from $x>0$ and $x<0$ valleys. This gives us the
condition of energy quantization, which turns out to be the equation
for the zeros of Airy's function
\[ Ai \Big( \frac{-E}{(2n_y + 1)^{2/3}} \Big) = 0. \]
This condition has a straightforward interpretation. The spectrum is
discrete, because the Airy's functions are analytic and therefore
have countable many zeros on the real axis. The few first
approximate eigenenergies are shown in table \ref{12 stanow dla boz}
on page \pageref{12 stanow dla boz}, together with some exact ones.
The compatibility is within $7 \%$. The assumption, that the
particle moves only in one of the valleys, breaks the space
symmetries of the
system, so the approximated energies will not have degeneracies.\\\\

The quantum supersymmetric hamiltonian has the form
            \begin{equation}
            \label{eq:h_susy_qm}
            \hat{H}_{susy} = \hat{p}^{2}_{x} + \hat{p}^{2}_{y} + \hat{x}^{2} \hat{y}^{2} + (\hat{x}+i\hat{y})f^{\dagger}
            +(\hat{x}-i\hat{y})f,
            \end{equation}
where $f^{\dagger}$ and $f$ are the fermionic creation and
annihilation operators respectively. The operator $f^{\dagger}f$
does not commute with the hamiltonian (\ref{eq:h_susy_qm}), so the
fermionic occupation number is not a good quantum number and cannot
be used to label the eigenstates. This generates non-vanishing,
non-diagonal matrix elements in the fermionic occupation number
representation.

    There exists one supersymmetry generator
            \begin{equation}
            \label{eq:susy_gen}
            \hat{Q} = \hat{Q}^{\dagger} = \frac{1}{\sqrt{2}} ( (\hat{p}_{x} +i \hat{p}_{y}) f + (\hat{p}_{x} -i \hat{p}_{y}) f^{\dagger} +   \hat{x}\hat{y} [ f^{\dagger},f ]
            ),
            \end{equation}
such that
            \[ \hat{H} = \{\hat{Q}^{\dagger}, \hat{Q} \}, \]
and
            \[ [ \hat{H}, \hat{Q} ] = 0. \]
    Although the supersymmetry concerns particles - bosons and fermions,
it is sometimes useful to think about the fermionic degree of
freedom as an equivalent spin projection on the OZ axis. In this
language the hamiltonian and the supersymmetry generator take the
form
            \[
            \hat{H}_{susy} = \hat{p}^{2}_{x} + \hat{p}^{2}_{y} + \hat{x}^{2} \hat{y}^{2} + \sigma_{x} \hat{x} - \sigma_{y}
            \hat{y},
            \]
            \[
            \hat{Q} = \hat{Q}^{\dagger} = \frac{1}{\sqrt{2}} (\hat{p}_{x} \sigma_{x} + \hat{p}_{y} \sigma_{y} +
            \hat{x}\hat{y}\sigma_z).
            \]
    The analytic analysis of the system (\ref{eq:h_susy_qm})
is difficult because of the mentioned non-diagonal matrix elements
in the fermionic occupation number representation. We can
nevertheless extract some useful information by analyzing a simpler
and more regular model. It will enable us to demonstrate the effect
of coexistence of the discrete and continuous spectrum in the same
range of energies. Let's consider the hamiltonian
            \begin{equation}
            \label{eq:h_susy_qm approx}
            \hat{H}_{model} = \hat{p}^{2}_{x} + \hat{p}^{2}_{y} + \hat{x}^{2} \hat{y}^{2} +
            \hat{x}[f^{\dagger},f].
            \end{equation}
We will use now, as in the purely bosonic case, the Born-Oppenheimer
approximation. It has to be noted that although the model
hamiltonian (\ref{eq:h_susy_qm approx}) does not have the space
symmetries present in the original system (\ref{eq:h_susy_qm}), it
does not spoil its usefulness, because the approximation breaks
these symmetries anyway.  Let's consider the motion in one of the
valleys, say $x>0$. The transverse potential is a potential of a
quantum supersymmetric harmonic oscillator of frequency $\omega \sim
|x|$. The hamiltonian of such oscillator is \cite{cooper}
            \[
            \hat{H}_{susy \ oscillator} = \frac{1}{2}\hat{p}^{2}_{y} + \frac{1}{2} |x|^{2} \hat{y}^{2}
            + \frac{|x|}{2}[f^{\dagger},f].
            \]
Its energies depend on one quantum number $n$ and are equal
            \[
            E_n = |x| n.
            \]
Similarly to the bosonic case, we write the wavefunction as a
product of two functions
\[ \Psi(x,y) = \alpha (x) \beta_{x,n} (y), \]
where $\alpha(x)$ accounts for the onward motion and
$\beta_{x,n}(y)$ fulfills the equation
            \[
            (-\frac{\partial^2}{\partial y^2} + x^2 y^2
            - |x|[f,f^{\dagger}]) \beta_{x,n}(y) = 2n|x|
\beta_{x,n}(y).
            \]
This gives us the equation on $\alpha (x)$
            \begin{equation}
            \label{eq:rownanie na alfe}
            (-\frac{\partial^2}{\partial x^2} + 2n|x|) \alpha_n(x) = E
\alpha_n(x).
            \end{equation}
From the form of (\ref{eq:rownanie na alfe}) we can conclude that
the spectrum has two coexisting parts: a continuous part for $n=0$
and a discrete one for $n \ge 1$. The zero energy of the zero-modes
enables the particle to penetrate the valley to any depth. The same
effect is present in the original \mbox{system
(\ref{eq:h_susy_qm})}, so it seems that (\ref{eq:h_susy_qm approx})
is a good candidate to model this feature.

A detailed study of spectra of both bosonic and supersymmetric
quantum models is described later on.

    \subsubsection*{Classical systems}

    The classical bosonic hamiltonian has the form
            \begin{equation}
            \label{eq:h_boz_cl}
            H_{boz} = p^{2}_{x} + p^{2}_{y} + x^{2}y^{2}.
            \end{equation}
We easily find the equations of motion
            \begin{eqnarray}
            \label{eq:eq_of_motion_boz}
            \dot{x} & = & 2 p_{x}, \nonumber \\
            \dot{y} & = & 2 p_{y}, \\
            \dot{p_{x}} & = & - 2 x y^{2}, \nonumber \\
            \dot{p_{y}} & = & - 2 y x^{2}. \nonumber
            \end{eqnarray}
These equations imply that particles which move exactly along the
coordinate axes have a constant momentum component along these axes.
This, of course, corresponds to a free motion in these directions.
We will show later that any other trajectory is bounded in the sense
that it always returns to the center of the potential. This property
of the $x^2y^2$ potential can be easily investigated using the
so-called hyperbolic billiard. The authors of the article
\cite{sieber} have considered a classical, free, spinless particle
moving as a free particle in a bounded set
\[D = \{(x,y)| x \ge 0 \land y \ge 0 \land y \le 1/x \},\] which is
an idealized version of the potential \mbox{in (\ref{eq:h_boz_cl})}.
They showed that such particle cannot escape through the valleys, it
always turns back. They also proved that it is possible to
systematically find all closed orbits. Every orbit can be
unambiguously labeled by a set of symbols which correspond to
reflection from particular walls of the potential. Such a list of
all orbits is particularly useful in the semiclassical calculations
by Feynman integrals.

\begin{figure}
        \begin{center}
        \includegraphics[width = 0.37\textwidth]{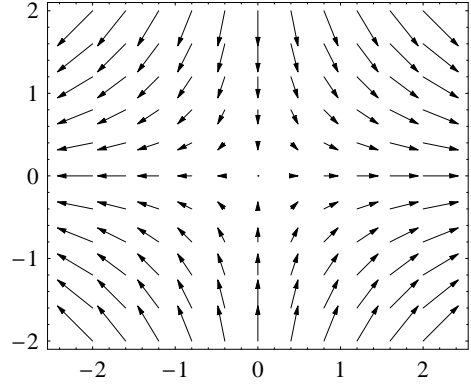}
        \caption{Vector field $\vec{V}$ about which the spin rotates. \label{vector_field}}
        \end{center}
        \end{figure}

    The classical equations of motion of the supersymmetric system
can be obtained from the quantum hamiltonian (\ref{eq:h_susy_qm})
using the Ehrenfest's equation for mean values
            \begin{eqnarray}
            \label{eq:eq_of_motion_susy}
            \dot{x} & = & 2 p_{x}, \nonumber \\
            \dot{y} & = & 2 p_{y}, \nonumber \\
            \dot{p_{x}} & = & - 2 x y^{2} - 2 S_{x}, \nonumber \\
            \dot{p_{y}} & = & - 2 y x^{2} + 2 S_{y}, \\
            \dot{S_{x}} & = & - 2 y S_{z}, \nonumber \\
            \dot{S_{y}} & = & - 2 x S_{z}, \nonumber \\
            \dot{S_{z}} & = & 2 x S_{y} + 2 y S_{x}. \nonumber
            \end{eqnarray}
Equations (\ref{eq:eq_of_motion_susy}) can also be derived in a
different way. As was shown in the paper by Berezin \cite{berezin},
for the description of classical dynamics of a nonrelativistic
particle with spin $\frac{1}{2}$ one needs an enriched phase space.
This description uses three, anticommuting, dynamical variables
$\xi_k$ which belong to a Grassmann algebra $G_3$ with three
generators.
\[ \xi_k \xi_l + \xi_l \xi_k = 0, \ k,l = 1,2,3 \]
In particular $ \xi^2_k = 0$. We can use relations between quantum
spin projection operators with their classical counterparts
expressed in terms of $\xi$. We thus obtain a classical form of the
supersymmetric hamiltonian
\[ H_{susy} = p^{2}_{x} + p^{2}_{y} + x^{2} y^{2} -i (x \epsilon_{xlm} - y
            \epsilon_{ylm}) \xi_l \xi_m. \]
The time dependence of the dynamical variables can be derived from
the Hamilton \mbox{equations :} \mbox{$ \frac{dq}{dt} =
\frac{\partial H}{\partial p}$}, \ \mbox{$\frac{dp}{dt} = -
\frac{\partial H}{\partial q}$}, \ \mbox{$\frac{d \xi_i}{dt} = i H
\frac{\overleftarrow{\partial}}{\partial \xi_i} \ $} where \mbox{$q
= x,y$} and \mbox{$p = p_x, p_y $}. The appearance of the imaginary
$i$ in the classical equations is inherently related to the nature
of the Grassmann variables.
\begin{eqnarray}
\label{eq:r ruchu susy z ksi}
\dot{x} & = & 2 p_x, \nonumber \\
\dot{y} & = & 2 p_y, \nonumber \\
\dot{p_x} & = & - 2xy^2 + i \epsilon_{xlm} \xi_l \xi_m, \nonumber \\
\dot{p_y} & = & - 2yx^2 - i \epsilon_{ylm} \xi_l \xi_m, \\
\dot{\xi_1} & = & - 2 y \xi_3, \nonumber \\
\dot{\xi_2} & = & - 2 x \xi_3, \nonumber \\
\dot{\xi_3} & = & 2 x \xi_2 + 2 y \xi_1. \nonumber
\end{eqnarray}
It is useful to replace the Grassmann variables $\xi_k$ in these
equations by the spin projections $S_k$, which are the physical
observables.
\[ S_k = -\frac{i}{2} \epsilon_{klm} \xi_l \xi_m, \]
\[ \dot{S}_k = -\frac{i}{2} \epsilon_{klm} (\dot{\xi}_l \xi_m +
\xi_l \dot{\xi}_m). \] We get rid of the time derivatives
$\dot{\xi_k}$ using the equations of motion (\ref{eq:r ruchu susy z
ksi}). The equations for $S_k$ close and we get the already obtained
set of equations (\ref{eq:eq_of_motion_susy}). In this language, the
hamiltonian takes the form
            \begin{equation}
            \label{eq:h_susy_cl}
            H_{susy} = p^{2}_{x} + p^{2}_{y} + x^{2} y^{2} + 2 x S_x - 2 y
            S_y.
            \end{equation}
Furthermore, equation (\ref{eq:h_susy_cl}) can be rewritten in a
form where the spin precession is evident. To this end we define a
vector field, which is space-dependent
\[ \vec{V} = ( V_x, V_y, V_z ) = ( 2x, -2y, 0). \]
Then
\[  H_{susy} = p^{2}_{x} + p^{2}_{y} + x^{2} y^{2} + \vec{S} \cdot \vec{V},
\] which manifestly shows the precession effect.
It is noteworthy that such a field is tangent to the contour line of
the potential $x^2y^2$, see figure \ref{vector_field}.

\subsection*{Quantum mechanics on PC} \label{sec:mechanika kwantowa
na PC}

    A numerical analysis of quantum systems can be easily carried out in
an eigenbasis of occupation number operators, a so-called Fock basis
\cite{wosiek}. The occupation number operator can be written as
$a^{\dagger}a$, where $a^{\dagger}$ i $a$ are standard bosonic
creation and annihilation operators respectively. They fulfill
well-known commutation relations
\begin{equation}
\label{eq:boz komutatory} [a_p, a_q] = [a_p^{\dagger},
a_q^{\dagger}] = 0,\quad [a_p, a_q^{\dagger}] = \delta_{pq}.
\end{equation}
$p$ and $q$ in (\ref{eq:boz komutatory}) are indices which label the
bosonic degrees of freedom. Particularly, the non-supersymmetric
system (\ref{eq:h_boz_qm}) has two bosonic degrees of freedom,
labeled $x$ i $y$. Thus, a basis state is described by two integers
\begin{equation}
|m,n> = \frac{1}{\sqrt{n!m!}}(a_x^{\dagger})^m (a_y^{\dagger})^n
|0>, \quad m,n \ge 0.
\end{equation}
Of course, in a numerical analysis it is impossible to use an
infinite basis, so a cut-off, $N_{cut}$, is needed, which limits the
number of basis states. There are many ways to introduce such a
cut-off. One of them, called a square cut-off, is to limit
independently the maximal number of occupation for each of two
degrees of freedom by $\sqrt{N_{cut}}$.

The momentum and position operators can be expressed by creation and
annihilation operators in the following way
\begin{eqnarray}
\label{eq:op x i y}
x = \frac{1}{\sqrt{2}}(a_x+a_x^{\dagger}) & , & \ p_x = \frac{1}{i\sqrt{2}}(a_x-a_x^{\dagger}), \\
y = \frac{1}{\sqrt{2}}(a_y+a_y^{\dagger}) & , & \ p_y =
\frac{1}{i\sqrt{2}}(a_y-a_y^{\dagger}). \nonumber
\end{eqnarray}

The action of the hamiltonian, which is an operator function of
(\ref{eq:op x i y}), is straightforward in the Fock basis. We can
easily calculate its matrix elements. The eigenenergies of a quantum
system belong to a set of eigenvalues of the hamiltonian matrix, and
the eigenstates are the eigenvectors of this matrix.

In the supersymmetric case we must introduce fermionic creation and
annihilation operators, $f$ and $f^{\dagger}$, which fulfill the
anticommutation relations
\begin{equation}
\label{eq:ferm komutatory} \{f_p, f_q\} = \{f_p^{\dagger},
f_q^{\dagger}\} = 0 , \quad \{f_p, f_q^{\dagger}\} = \delta_{pq},
\end{equation}
where $p$ and $q$ are indices which describe the fermionic degrees
of freedom. In general, in order to ensure the relations
(\ref{eq:ferm komutatory}) one uses a construction by Jordan and
Wigner \cite{wigner}. In a case of only one fermionic degree of
freedom it is not necessary. The fermionic occupation number operator
$f^{\dagger}f$, with two eigenvalues $0$ i $1$, permit to label the
basis states by a third quantum number
\[
|m,n,k> = \frac{1}{\sqrt{n!m!}}(a_x^{\dagger})^m (a_y^{\dagger})^n
(f^{\dagger})^k |0>, \]
\begin{equation}
m,n \ge 0, k=0,1.
\end{equation}
\par
\begin{figure}[!t]
        \begin{center}
        \includegraphics{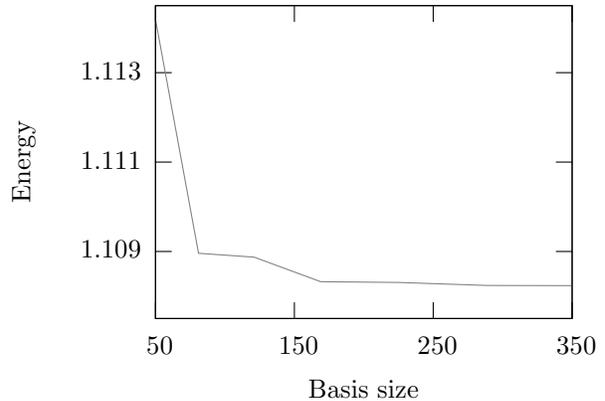}
        \caption{The dependence of the ground state energy on the cut-off for the bosonic system. \label{fit_boz}}
        \end{center}
        \end{figure}

The above method allows also to verify the reliability of numerical
results. The rate of convergence of eigenenergies with an increasing
$N_{cut}$ is a simple criterion. For example, we can assume that the
convergence is reached when a relative change of the energy between
consecutive $N_{cut}$ is less than $1\%$ of its absolute value. For
the bosonic ground state this happens for $N_{cut} > 100$. The
excited states have a worse convergence so all the following results
were obtained for $N_{cut} = 400$. The dependence of the energy of
the bosonic ground state on the cut-off is shown in figure
\ref{fit_boz} with a solid line.
        \begin{figure}[!t]
        \begin{center}
        \includegraphics{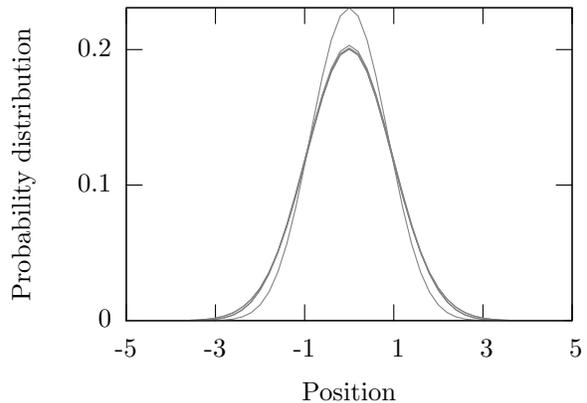}
        \caption{The dependence of the space distribution of the bosonic ground state on the cut-off $N_{cut}$, for $N_{cut}$ equal 16 (highest curve), 36, 64, 128 (lowest curve). \label{fit_boz_func}}
        \end{center}
        \end{figure}

This dependence provides us also with an additional information on
the character of the quantum state. Particularly, as it will be
described later, it enables one to decide whether the state belongs
to the discrete spectrum or to the continuous one. Another criterion
of reliability is supplied by the spatial probability distributions.
One can assume that the convergence is reached when for consecutive
cut-offs, $N_{cut}$ and $N'_{cut}$, the probability $P(x)$ of
finding the particle at the point $x$ fulfill the inequality
\[ \max_{ \ x \ \in \ I} \frac{| P_{N_{cut}}(x) - P_{N'_{cut}}(x) |}{P_{N_{cut}}(x)}
< \epsilon, \] where $I$ is an interval on which the distributions
are calculated. Figure \ref{fit_boz_func} shows a cross-section
along the OX axis of such distribution for the bosonic ground state
for different cut-offs. The convergence is reached for $N_{cut} =
128$. The free parameters were chosen as $I = [ -5, 5]$ and
$\epsilon = 0.01$. Obviously these values are only exemplary and can
be tuned at will.

\subsection*{Symmetries of the systems} \label{sec:symetrie ukladow}

\subsubsection*{Bosonic case and the $C_{4v}$ group}

    The hamiltonian of the quantum bosonic \mbox{system (\ref{eq:h_boz_qm})} is
invariant with respect to a point group of symmetries, which
consists of a 4-fold principal axis of symmetry and four reflection
axes perpendicular to the principal axis, and is called the $C_{4v}$ group. It has eight elements\\
\begin{tabular}{rcl}
$e$ & - & identity, \\
$\pi_{x} $ & - & reflection \ with \ respect \\
& & \ to \ the \ $x$ \ axis , \\
$\pi_{y} $ & - & reflection  \ with \ respect \\
& & \ to \ the \ $y$ \ axis , \\
$\pi_{x=y} $ & - & reflection \ with \ respect \\
& & \ to \ the \ $x=y$ \ line , \\
$\pi_{x=-y} $ & - & reflection \ with \ respect \\
& & \ to \ the \ $x=-y$ \ line , \\
$R_{\frac{\pi}{2}} \equiv R $ & - & rotation \ through \\
& & \ the \ angle \ of \ $+ \frac{\pi}{2}$ \ radians, \\
$R_{\pi} \equiv R^2 $ & - & rotation \ through \\
& & \ the \ angle \ of \ $+ \pi$ \ radians, \\
$R_{\frac{3\pi}{2}} \equiv R^3 $ & - & rotation \ through \\
& & \ the \ angle \ of \ $ + \frac{3\pi}{2}$ \ radians.
\end{tabular}
\\
\\
We call a conjugacy class a subset of all elements of a group G that
commute with all elements of G. The group $C_{4v}$ has five
conjugacy classes
\begin{eqnarray}
id & = & \{ e \}, \nonumber \\
2C_4 & = & \{ R , R^3 \}, \nonumber \\
C_2 & = & \{ R^2 \}, \nonumber \\
2S_v & = & \{ \pi_x , \pi_y \}, \nonumber \\
2S_d & = & \{ \pi_{x=y} , \pi_{x=-y} \}. \nonumber
\end{eqnarray}
It is known \cite{georgi}, that the number of irreducibles
representations of a group is equal to the number of classes of this
group. At the same time, the sum of squares of dimensions of this
representations must be equal to the order of the group. All this
implies that $C_{4v}$ must have four 1-dimensional and one
2-dimensional irreducible representations. The defining
representation is \mbox{2-dimensional} and is explicitly shown in
(\ref{rep_2d}). Each group has also a 1-dimensional, trivial
representation composed of ones.
\begin{eqnarray}
\label{rep_2d}
e = \left(\begin{array}{cc}
1&0 \\
0&1 \\
\end{array}\right) , &
R = \left(\begin{array}{cc}
0&1 \\
-1&0 \\
\end{array}\right) ,\nonumber \\
R^2 = \left(\begin{array}{cc}
-1&0 \\
0&-1 \\
\end{array}\right) , &
R^3 = \left(\begin{array}{cc}
0&-1 \\
1&0 \\
\end{array}\right) ,\nonumber \\
\pi_x = \left(\begin{array}{cc}
-1&0 \\
0&1 \\
\end{array}\right) , &
\pi_y = \left(\begin{array}{cc}
1&0 \\
0&-1 \\
\end{array}\right) ,\\
\pi_{x=y} = \left(\begin{array}{cc}
0&1 \\
1&0 \\
\end{array}\right) , &
\pi_{x=-y} = \left(\begin{array}{cc}
0&-1 \\
-1&0 \\
\end{array}\right). \nonumber &
\end{eqnarray}
In order to find the remaining three 1-dimensional representations
one uses a couple of observations. First, every representation of
the identity element has to be equal 1. Second, except the elements
of the $2C_4$ class, all elements squared give the identity element.
Thus, their representations must be equal $+1$ or $-1$. Moreover,
one can obtain the elements of the $C_2$ class by combining the
elements of the classes $2C_4$, $2S_v$ and $2S_d$ (i.e. reflection
with respect to the OX axis and a reflection with respect to the OY
axis give a rotation through an angle of $\pi$). So the
representations of the $C_2$ class must be equal to $+1$. As the
representations of the $2C_4$ class are concerned, they can be equal
$+1$ or $-1$, because when squared they must give the elements of
the $C_2$ class. In order that the multiplication table remains
unchanged the representations equal to $-1$ must appear in pairs.
All these observations permit to identify all 1-dimensional
representations of the $C_{4v}$ group (table \ref{rep_1d}).
\begin{table}[!ht]
\begin{center}
\begin{tabular}{|c|c|c|c|c|c|}
\hline
    & e & $2C_4$& $C_2$ & $2S_v$ & $2S_d$ \\
\hline
$A_1$   & 1 & 1 & 1 & 1 & 1 \\
\hline
$A_2$   & 1 & 1 & 1 & -1    & -1    \\
\hline
$B_1$   & 1 & -1    & 1 & 1 & -1    \\
\hline
$B_2$   & 1 & -1    & 1 & -1    & 1 \\
\hline
\end{tabular}
\caption{Four 1-dimensional irreducible representations of the
$C_{4v}$ group. \label{rep_1d}}
\end{center}
\end{table}
The eigenstates transform according to one of the irreducible
representation. We have four 1-dimensional representations, $A_1,
A_2, B_1$ and $B_2$, so we will have four series of nondegenerate
states. States transforming according to the representation $E$ will
be doubly degenerate. A state symmetric with respect to three
reflections $\pi_x$, $\pi_y$ and $\pi_{x=y}$, will transform
according to the trivial representation $A_1$. A state symmetric
with respect to $\pi_x$ and $\pi_y$ reflections but antisymmetric
with respect to $\pi_{x=y}$ will transform according to the $B_1$
representations. \mbox{Table \ref{12 stanow dla boz}} contains first
twelve states, their symmetries and the representation they belong
to.

\subsubsection*{Supersymmetric case}

    The quantum supersymmetric hamiltonian remains unchanged under
action of transformations which form the following group

\begin{tabular}{rccccc}
$e                  $ & - & identity ,& & & \\
$\tilde{\pi}_x       $ & - & $x \rightarrow -x,        $&$ y \rightarrow \quad y,$ \\
& &$ f \rightarrow -f^{\dagger},     $&$ f^{\dagger} \rightarrow -f$, \\
$\tilde{\pi}_y       $ & - & $x \rightarrow \quad x,   $&$ y \rightarrow -y,$ \\
& &$ f \rightarrow \quad f^{\dagger},$&$ f^{\dagger} \rightarrow \quad f$, \\
$\tilde{R}_{\pi}    $ & - & $x \rightarrow -x,        $&$ y \rightarrow -y,       $\\
& &$ f \rightarrow -f,               $&$ f^{\dagger} \rightarrow -f^{\dagger}$, \\
$\tilde{\pi}_{x=y}  $ & - & $x \rightarrow \quad y,   $&$ y \rightarrow \quad x,  $\\
& &$ f \rightarrow if^{\dagger},     $&$ f^{\dagger} \rightarrow -if$, \\
$\tilde{\pi}_{x=-y} $ & - & $x \rightarrow -y,        $&$ y \rightarrow -x,       $\\
& &$ f \rightarrow -if^{\dagger},    $&$ f^{\dagger} \rightarrow if$, \\
$\tilde{R}_{\frac{\pi}{2}}   $ & - & $x \rightarrow -y,        $&$ y \rightarrow \quad x,  $\\
& &$ f \rightarrow if,               $&$ f^{\dagger} \rightarrow -if^{\dagger}$, \\
$\tilde{R}_{\frac{3\pi}{2}}     $ & - & $x \rightarrow \quad y, $&$
y \rightarrow -x, $\\
& &$ f \rightarrow -if, $&$ f^{\dagger}\rightarrow if^{\dagger}$.
\end{tabular}
\\
\\

\begin{table*}
\begin{center}
\begin{tabular}{|c|c|c|c|c|c|c|c|c|c|c|}
\hline
L.p. & Energy & B-O & $\pi_x$ & $\pi_y$ & $\pi_{x=y}$ & $\pi_{x=-y}$ & $R$ & $R^2$ & $R^3$ & Represen-\\
&  &  & & & & & & & & tation \\

\hline
1 & 1.1082   &  1.1737              & + & + & +   & +    & + & +   & +   & $A_1$ \\
2 & 2.3788   &  2.3381              & - & + & 0   & 0    & 0 & -   & 0   & $E$ \\
3 & 2.3788   &  2.4414              & + & - & 0   & 0    & 0 & -   & 0   & $E$ \\
4 & 3.0574   &  3.2711              & + & + & -   & -    & - & +   & -   & $B_1$ \\
5 & 3.5229   &  3.4319              & + & + & +   & +    & + & +   & +   & $A_1$ \\
6 & 4.1100   &  4.0878              & + & - & 0   & 0    & 0 & -   & 0   & $E$ \\
7 & 4.1100   &  4.2950              & - & + & 0   & 0    & 0 & -   & 0   & $E$ \\
8 & 4.8210   &  4.8307              & + & + & -   & -    & - & +   & -   & $B_1$ \\
9 & 5.0113   &  4.8635              & - & - & +   & +    & - & +   & -   & $B_2$ \\
10 & 5.1120  &  5.0783              & + & + & +   & +    & + & +   & +   & $A_1$ \\
11 & 5.6947  &  5.5206              & + & - & 0   & 0    & 0 & -   & 0   & $E$ \\
12 & 5.6947  &  5.8053              & - & + & 0   & 0    & 0 & -   & 0   & $E$ \\
\hline
\end{tabular}
\caption{First twelve eigenstates of the quantum bosonic system with
a corresponding representation of the symmetry group $C_{4v}$. In
the column entitled 'Energy' the exact energies are shown, whereas
the 'B-O' column contains values calculated in the Born-Oppenheimer
approximation. In the columns from 4 to 10 are shown the parity or a
lack of a given symmetry for each state. \label{12 stanow dla boz}}
\end{center}
\end{table*}

The aim of this notation is to highlight the similarity of the above
group and the $C_{4v}$ group. In this spirit we call this group the
$\tilde{C}_{4v}$ group. We again find five conjugacy classes
\begin{eqnarray}
id & = & \{ e \}, \nonumber \\
2\tilde{C}_4 & = & \{ \tilde{R}_{\frac{\pi}{2}} , \tilde{R}_{\frac{3\pi}{2}} \}, \nonumber \\
\tilde{C}_2 & = & \{ \tilde{R}_{\pi} \}, \nonumber \\
2\tilde{S}_v & = & \{ \tilde{\pi}_x , \tilde{\pi}_y \}, \nonumber \\
2\tilde{S}_d & = & \{ \tilde{\pi}_{x=y} , \tilde{\pi}_{x=-y} \}.
\nonumber
\end{eqnarray}
We can perform a similar analysis as in the bosonic case and obtain
four \mbox{1-dimensional} and one 2-dimensional irreducible representations
(table \ref{rep_1d_susy} and formula \ref{rep_2d_susy}).
\begin{table}[!ht]
\begin{center}
\begin{tabular}{|c|c|c|c|c|c|}
\hline
    & & & & & \\
    & e & $2\tilde{S}_v$ & $\tilde{C}_2$ & $2\tilde{C}_4$ & $2\tilde{S}_d$ \\
\hline
$A$   & 1 & 1  & 1 & 1  & 1 \\
\hline
$B$   & 1 & -1 & 1 & 1  & -1\\
\hline
$C$   & 1 & 1  & 1 & -1 & -1\\
\hline
$D$   & 1 & -1 & 1 & -1 & 1 \\
\hline
\end{tabular}
\caption{Four 1-dimensional irreducible representations. \label{rep_1d_susy}}
\end{center}
\end{table}
\begin{eqnarray}
\label{rep_2d_susy}
e = \left(\begin{array}{cccc}
1&0 \\
0&1 \\
\end{array}\right) , &
\tilde{\pi}_x = \left(\begin{array}{cc}
-1&0 \\
0&1 \\
\end{array}\right) , \nonumber \\
\tilde{\pi}_y = \left(\begin{array}{cc}
1&0 \\
0&-1 \\
\end{array}\right), &
\tilde{R}_{\pi} = \left(\begin{array}{cc}
-1&0 \\
0&-1 \\
\end{array}\right) , \nonumber \\
\tilde{\pi}_{x=y} = \left(\begin{array}{cc}
0&1 \\
1&0 \\
\end{array}\right) , &
\tilde{\pi}_{x=-y} = \left(\begin{array}{cc}
0&-1 \\
-1&0 \\
\end{array}\right), \nonumber \\
\tilde{R}_{\frac{\pi}{2}} = \left(\begin{array}{cc}
0&-1 \\
1&0 \\
\end{array}\right) , &
\tilde{R}_{\frac{3\pi}{2}} = \left(\begin{array}{cc}
0&1 \\
-1&0 \\
\end{array}\right). &
\end{eqnarray}
Since all elements of the $\tilde{C}_{4v}$ group commute with the hamiltonian,
one can choose one of these symmetries in order to define an additional
quantum number. Let's choose the following
\[\tilde{\pi}_y : x \rightarrow x, \quad y \rightarrow -y, \quad f \rightarrow f^{\dagger}, \quad f^{\dagger} \rightarrow f. \]
The action of $\tilde{\pi}_y$ on basis vectors is given by
\[\tilde{\pi}_y | n_x, n_y, n_f > = (-)^{n_y} |n_x, n_y, 1 - n_f>. \]
One can define even and odd states with respect to $\tilde{\pi}_y$
\[ |n_x, n_y,\pm > = \frac{1}{\sqrt{2}}(|n_x, n_y, 0> \pm
(-)^{n_y}|n_x, n_y, 1>.\] In such a basis the hamiltonian appears as
a block diagonal matrix with two sectors, and it is possible to
diagonalize each block independently. Although we have introduced a
squared cut-off the basis remains invariant with respect to the
symmetries from the $\tilde{C}_{4v}$ group. The spectrum of the
supersymmetric system (\ref{eq:h_susy_qm}) will have degeneracies
due to these symmetries for each cut-off. It turns out that all
states transform according to the 2-dimensional irreducible
representation, so they will all be doubly degenerate. Table
\ref{energie_susy} shows
eight first energies for three different cut-offs.\\
\begin{table}[!ht]
\begin{center}
\begin{tabular}{|c|c|c|c|}
\hline
    & $N_{cut}=72 $& $N_{cut}=128 $ & $N_{cut}=200 $ \\
\hline
1 & 0.2871 & 0.2469 & 0.1788 \\
2 & 0.2871 & 0.2469 & 0.1788 \\
\hline
3 & 1.2172 & 0.8573 & 0.7553 \\
4 & 1.2172 & 0.8573 & 0.7553 \\
\hline
5 & 2.2215 & 1.8932 & 1.5014 \\
6 & 2.2215 & 1.8932 & 1.5014 \\
\hline
7 & 3.1867 & 2.7090 & 2.4616 \\
8 & 3.1867 & 2.7090 & 2.4616 \\
\hline
\end{tabular}
\caption{Eight first energies for three different cut-offs for the
supersymmetric case. The degeneracy due to the point symmetries
is exact for each cut-off. \label{energie_susy} }
\end{center}
\end{table}

\textbf{Supersymmetry} \\

   Supersymmetry can manifest itself in a given quantum system
through the set of supersymmetry generators, which form a
well-defined algebra, or, on a more experimental level, through a
characteristic structure (degeneracies) of the spectrum. In this
paper we assume that the existence of a single ground state with
zero energy and supersymmetric doublets of higher energies
guarantees the presence of supersymmetry. Indeed, in the case of
\mbox{system (\ref{eq:h_susy_qm})} the generators $Q$ and
$Q^{\dagger}$ do not fulfill all relations of a supersymmetry
algebra, namely $Q^2 \ne 0$. Moreover, $Q$ does not conserve the
bosonic occupation number, so the introduction of the cut-off breaks
the supersymmetry and destroys its fingerprints on the spectrum. One
expects its restoration in the limit \mbox{$ N_{cut} \rightarrow
\infty $}. However, there exist a way of establishing the
supersymmetry for each finite $N_{cut}$. One can remark, that, since
$Q$ is hermitian, the hamiltonian (\ref{eq:h_susy_qm}) is a product
of generators $Q$ specified for an infinite cut-off, and then cut to
the desired dimensions. On the other hand, one can conceive the
hamiltonian matrix as a product of matrices of already cut
generators $Q$.
    \begin{equation}
    \label{eq:h_qq}
    [ H_{susy} ] = [Q_{finite \ N_{cut} }][ Q_{finite \ N_{cut}}].
    \end{equation}
Of course, the hamiltonians (\ref{eq:h_susy_qm}) and (\ref{eq:h_qq})
agree in the limit of infinite cut-off. It turns out that the
spectrum of (\ref{eq:h_qq}) is fully supersymmetric. A sample of
results for three different cut-offs are shown in table
\ref{energie_qq}. One can observe a double degeneracy due to the
point symmetries of the system, as well as a supersymmetric
degeneracy.
\begin{table}[!ht]
\begin{center}
\begin{tabular}{|c|c|c|c|}
\hline
  $N_{cut}=50 $& $N_{cut}=200 $ & $N_{cut}=450 $ \\
\hline
0.0000 & 0.0165 & 0.0000 \\
0.0000 & 0.0165 & 0.0000 \\
       & 0.0165 &        \\
       & 0.0165 &        \\
\hline
0.3371 & 0.3607 & 0.0862 \\
0.3371 & 0.3607 & 0.0862 \\
0.3371 & 0.3607 & 0.0862 \\
0.3371 & 0.3607 & 0.0862 \\
\hline
1.0163 & 0.8867 & 0.3760 \\
1.0163 & 0.8867 & 0.3760 \\
1.0163 & 0.8867 & 0.3760 \\
1.0163 & 0.8867 & 0.3760 \\
\hline
\end{tabular}
\caption{Twelve first eigenvalues of the matrix $[H_{susy}]$ for
three different cut-offs. The degeneracies have two origins:
supersymmetry and point symmetries of the group $\tilde{C}_{4v}$.
For cut-off, for which $N_{cut}/2$ is even, the supersymmetry is
broken and the spectrum does not have the zero energy ground state.
\label{energie_qq} }
\end{center}
\end{table}
Moreover, one can notice in table \ref{energie_qq} yet another way
of supersymmetry breaking. For square cut-offs with $N_{cut}/2$
even, the spectrum can not contain a single ground state and
$N_{cut}/4 -1$ supersymmetric doublets. In such situation the
supersymmetry is broken and the ground state disappears.

\subsection*{Quantum systems} \label{sec:analiza ukladow kwantowych}

    In this part we will compare the numerical spectra of the
bosonic and supersymmetric quantum system.\\

\begin{figure}[!t]
        \begin{center}
        \includegraphics{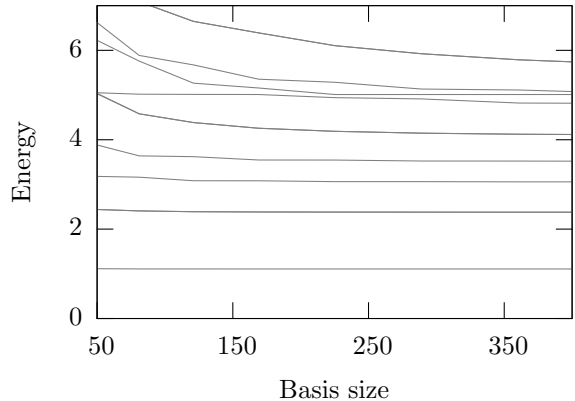}
        \caption{The dependence of the twelve first eigenenergies on the cut-off for the bosonic case. \label{zal_boz} }
        \end{center}
        \end{figure}

\begin{figure}[!t]
        \begin{center}
        \includegraphics{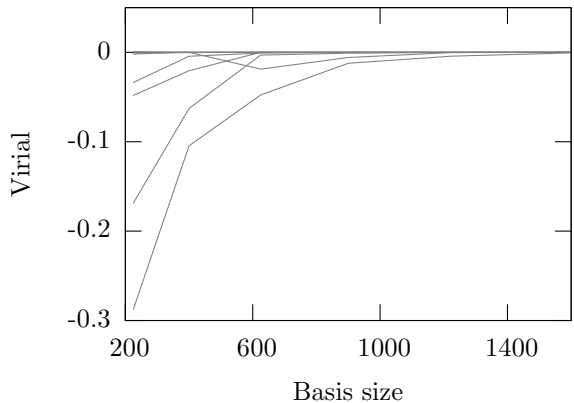}
        \caption{The dependence of the quantum virial on cut-off for ten first bosonic states.
        The convergence for the ground state is the fastest. The most
        distant from zero curve corresponds to the tenth eigenstate. \label{wir_qm_boson} }
        \end{center}
        \end{figure}

A part of the spectrum of the bosonic system (\ref{eq:h_boz_qm}) is
shown in \mbox{figure \ref{zal_boz}}. The dependence of twelve first
eigenenergies on the cut-off is depicted. Numerical values are
presented in table \ref{12 stanow dla boz} on page \pageref{12
stanow dla boz}.
Consequently to the discussion of the symmetries of the system, the
spectrum contains four nondegenerate series of states and one doubly
degenerate. The degeneracies due to the point symmetries are present
for each $N_{cut}$, because the cut-off does not spoil them.
        \begin{figure}[!t]
        \begin{center}
        \includegraphics{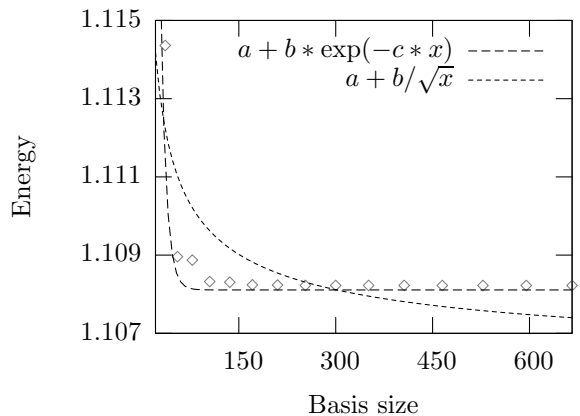}
        \caption{The dependence of the bosonic ground state energy on the cut-off with fitted
curves: an exponential one and an inverse square root.
\label{boz_fit} }
        \end{center}
        \end{figure}
    After the paper \cite{trzetrzelewski} the dependence of
the eigenenergy on the cut-off enables one to decide whether the
state belongs to the discrete spectrum or to the continuous one. If
the dependence on $N_{cut}$ is of the type $\frac{1}{N_{cut}}$ or
slower then the corresponding state is a nonlocalized state from the
continuous spectrum. Contrary, if the dependence is fast, for
example exponential $\sim \ e^{-N_{cut}}$, then the state is
localized. Figure \ref{boz_fit} shows the dependence of the bosonic
ground state on the cut-off and two fitted curves: an exponent
 $f(N_{cut}) = a + b \ e^{-c \ N_{cut}}$ and an inverse square root $g(N_{cut}) = a+b/\sqrt{N_{cut}}$.
The first one fits much better which means that the ground state is
localized. It turns out that all other states analyzed with this
method belong also to the discrete spectrum. This is also confirmed
by an analysis of the virial, which is a scalar product of vectors
of momentum and position $ w = \vec{p} \cdot \vec{x} $. In the
classical regime, the mean time derivative of a virial tends to zero
on a bounded trajectory and explodes on an unbounded one. On the
quantum level, the evaluation of the expectation value of a
corresponding operator in a given state can supply similar type of
information. In the bosonic case one is left with the following
expression
        \begin{equation}
        \label{eq:wirial_boz}
        <\dot{w}_{bosonic}> = 2 (\langle \hat{p}_{x}^{2}
\rangle + \langle \hat{p}_{y}^{2} \rangle - 2\langle
\hat{x}^{2}\hat{y}^{2} \rangle ).
        \end{equation}
The results for ten first states are shown in figure
\ref{wir_qm_boson} as a function of cut-off. We notice that all
values converge to zero, which confirms that the bosonic states are
localized.
\par
\begin{figure}[!t]
        \begin{center}
        \includegraphics{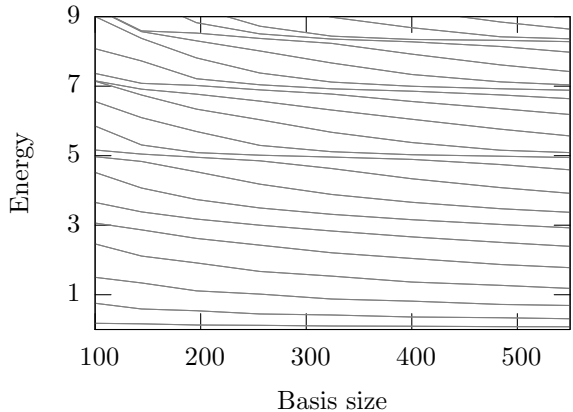}
        \caption{The dependence of the thirty two first states on the cut-off for the supersymmetric case. \label{zal_susy} }
        \end{center}
        \end{figure}
\begin{figure}[!t]
        \begin{center}
        \includegraphics{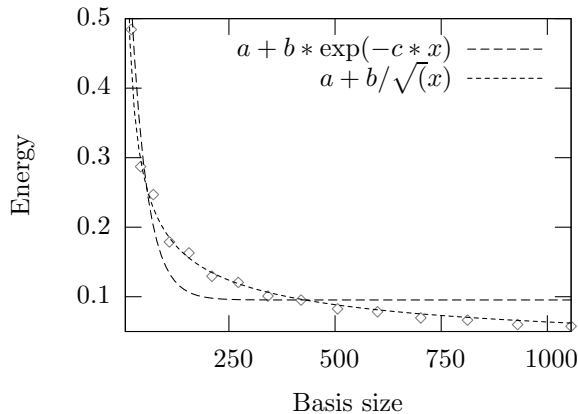}
        \caption{The dependence of the supersymmetric ground state energy on the cut-off with fitted
curves: an exponential one and an inverse square root.
\label{susy_fit} }
        \end{center}
        \end{figure}

On the other hand, the supersymmetric spectrum is shown in figure
\ref{zal_susy}. According to the symmetries of the system, all
states are doubly degenerate due to the point symmetries from the
$\tilde{C}_{4v}$ group. The supersymmetric degeneracy can be found
in the infinite cut-off limit.
The fitted cut-off dependence of the supersymmetric ground state is
shown in figure \ref{susy_fit} and suggest that it is a nonlocalized
state. The analysis of higher states also confirms that they belong
to the continuous spectrum.
Thus, it seems that the spectrum of the quantum supersymmetric
system is composed of purely nonlocalized states. But, the
approximate solution of the supersymmetric model (\ref{eq:h_susy_qm
approx}) indicates that among them one might find localized states
from the discrete spectrum. And indeed we have found such states in
the numerical spectrum of the exact hamiltonian
(\ref{eq:h_susy_qm}). However, as their convergence is exponential
we may expect that they should cross states with slower cut-off
dependence. The rules of quantum mechanics claim that when a
parameter in a hamiltonian is changed, the states of the same
symmetry cannot cross. In our case, such a parameter is our cut-off,
and as was mentioned, all states transform according to the
2-dimensional irreducible representation of the $\tilde{C}_{4v}$
group. So these localized states must have an interesting
realization. One can see them in figure \ref{zal_susy} as
deformations of the energy dependence on the cut-off. More
precisely, there exist states which on some interval $N_{cut}$ have
a constant energy (i.e. the deformations for energies around $E =
5$). If the energy of such a state remained unchanged till some
cut-off $\tilde{N}_{cut}$, then for $N_{cut} > \tilde{N}_{cut}$ it
starts decreasing as $\frac{1}{N_{cut}}$. However, from
$\tilde{N}_{cut}$ the energy of the higher, neighboring state
assumes this value (here $E=5$) and remains constant on some
adjacent interval. In this way localized states coexist with
nonlocalized ones in the same range of energy.
        \begin{figure}[!t]
        \begin{center}
        \includegraphics{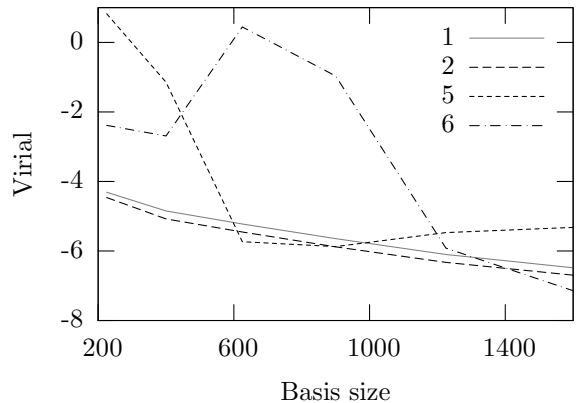}
        \caption{The dependence of the quantum virial on the cut-off for some selected
supersymmetric states (first, second, fifth and
sixth).\label{wir_qm_susy}}
        \end{center}
        \end{figure}
These results can again be confirmed by the virial analysis. We have
to evaluate the operator
        \begin{equation}
        \label{eq:wirial_susy}
        <\dot{w}_{susy}> = 2\langle \hat{p}_{x}^{2} \rangle + 2\langle \hat{p}_{y}^{2} \rangle -
4\langle {\hat{x}}^{2} {\hat{y}}^{2} \rangle - \langle \sigma_{x}
\hat{x} \rangle + \langle  \sigma_{y} \hat{y} \rangle.
        \end{equation}
        \begin{figure*}[!t]
        \begin{center}
        \includegraphics{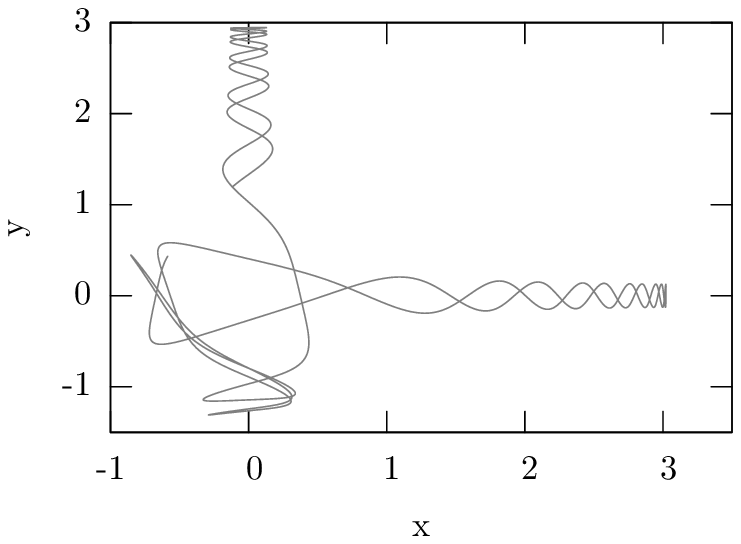}
        \includegraphics{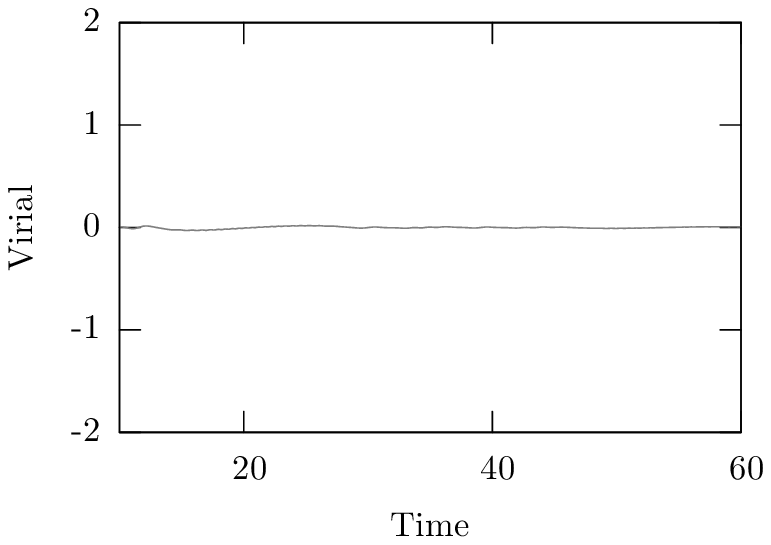}
        \caption{An exemplary bosonic classical trajectory and the virial
        dependence on time. The mean time derivative of the virial has
        small values indicating that the trajectory is bounded. \label{wirial_klasyczny_bosonowy}}
        \end{center}
        \end{figure*}
In figure \ref{wir_qm_susy} the time derivative of virial for four
supersymmetric states is depicted as a function of the cut-off. For
a majority of them the obtained values are far from 0. These states
are nonlocalized. However, as was observed there can exist also some
localized states. Taking into account their realization through many
nonlocalized states, one concludes that the states for which the
quantum virial has a value near 0 for a given cut-off, are just the
states that form the localized state.
\par
To summarize, the main observation is that bosonic and
supersymmetric eigenstates have a different character. The first one
are localized and thus belong to the discrete spectrum, whereas the
supersymmetric spectrum is composed from both types of states. This
disparity is essential in the following discussion.

\subsection*{Classical systems} \label{sec:analiza ukladow
klasycznych}
\par
    The differences in the behavior of classical bosonic and supersymmetric
trajectories are striking at the first glance. Similarly as in
quantum regime the bosonic states were localized and the
supersymmetric were nonlocalized, the bosonic trajectories are
bounded and the supersymmetric escape through the flat directions of
the potential.

\begin{figure*}[!ht]
        \begin{center}
        \includegraphics{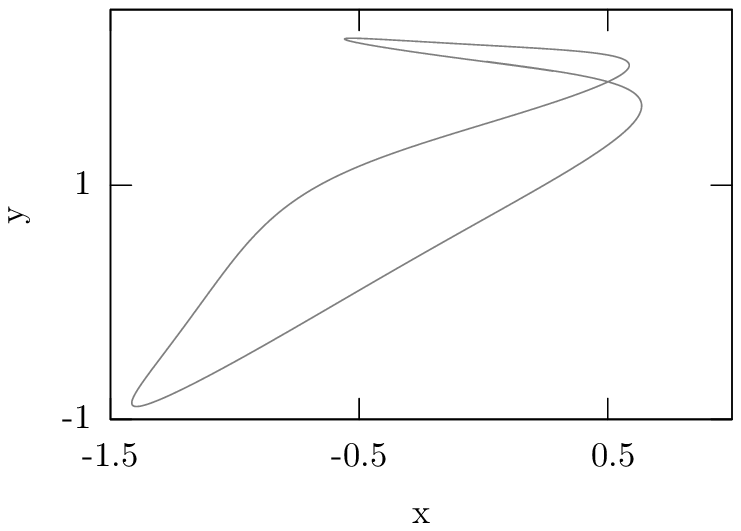}
        \includegraphics{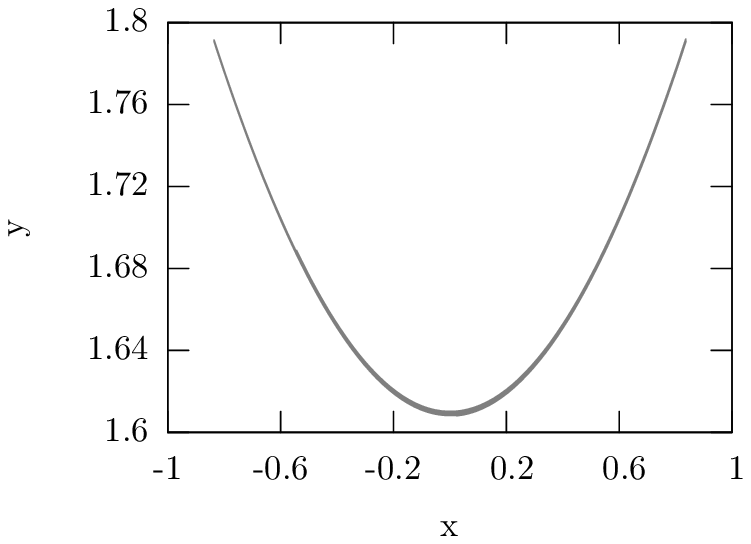}
        \caption{Closed bosonic and supersymmetric orbits. \label{bosonowe trajektorie zamkniete}}
        \end{center}
        \end{figure*}

    More precisely, the numerical solutions of bosonic equations of motions
(\ref{eq:eq_of_motion_boz}) show that all investigated trajectories
are bounded. They all come back to the center of the potential.
Similarly, as in the case of the hyperbolic billiard potential,
mentioned earlier, only particles that move along the coordinate
axes can escape through the valleys. In the quantum regime, we used
the virial theorem to decide whether quantum states were localized
or not. Now, we would like to do the same using the classical virial
theorem. It implies that for a bounded motion (with finite momentum
and position) in a potential that vanishes as $\frac{1}{r}$ one has
        \begin{equation}
        \label{eq:wirial_cl}
        0 = \overline{ \frac{dw}{dt} } = \overline{U} + 2 \overline{K},
        \end{equation}
where $w$ is the virial, $\frac{d}{dt}$ is the time derivative, ---
stands for the time averaging, U is the potential energy and K is
the kinetic energy. In order to verify the character of the
classical trajectories, one needs to evaluate the quantity
(\ref{eq:wirial_cl}) on a given trajectory. A representative bosonic
trajectory and the corresponding plot of the virial are shown in
figure \ref{wirial_klasyczny_bosonowy}. The mean time derivative of
the virial remains near zero, which indicates that the trajectory is
bounded. Moreover, using the algorithm proposed by Schmelcher and
Diakonos \cite{pingel} we found some closed orbits, one of which is
shown in figure \ref{bosonowe trajektorie zamkniete}. Investigation
of eigenvalues of the Jacobian shows that this orbit is unstable. It
is worth noting that similar research was made by Dalhqvist and
Russberg \cite{dahlqvist} who found a stable island in the phase
space of the $x^2y^2$ potential, therefore showing that the latter
is not ergodic as was believed.

\begin{figure*}[!ht]
        \begin{center}
        \includegraphics{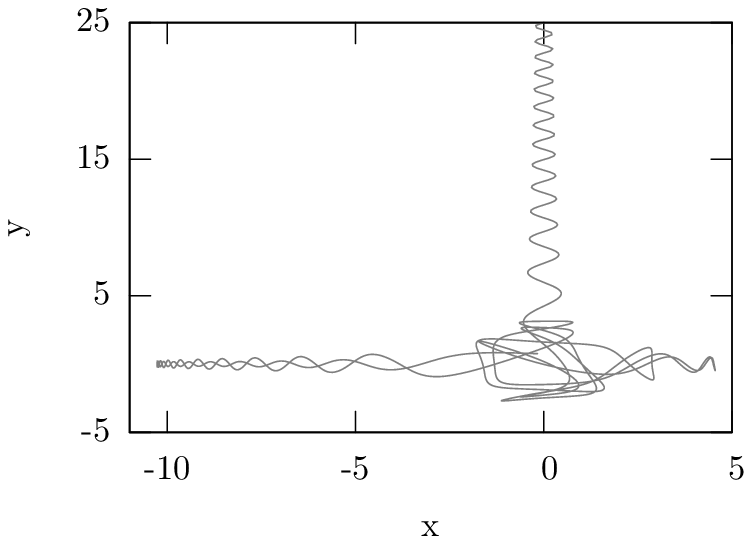}
        \includegraphics{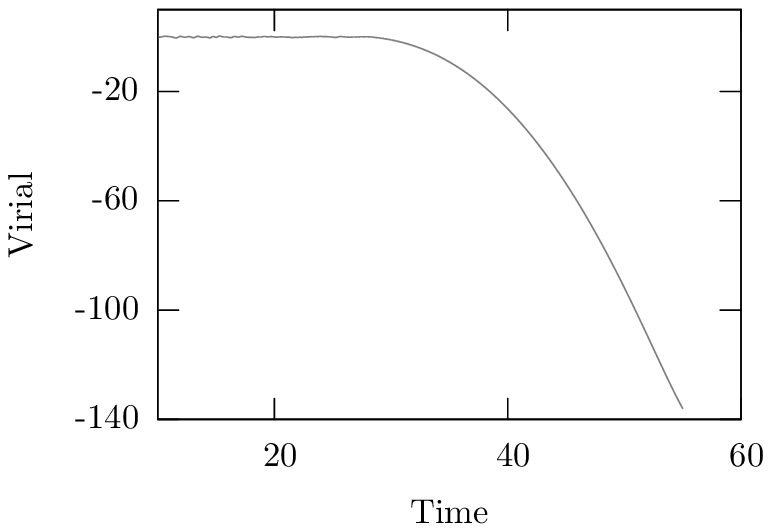}
        \caption{An exemplary supersymmetric classical trajectory and the virial
        dependence on time. The mean time derivative of the virial
        explodes when the particle escapes through the valley. \label{wirial_klasyczny_susy}}
        \end{center}
        \end{figure*}

On the contrary, all supersymmetric trajectories escape through the
valleys of potential. However, there exist some of them which for a
large amount of time remain in the center of potential. When the
dependence of the number of returns of some trajectory on its energy
is investigated, one finds sharp peaks suggesting that there exist
trajectories with a huge number of returns. Again, we found one
unstable closed orbit using the Schmelcher's algorithm, shown in
figure \ref{bosonowe trajektorie zamkniete}. The virial theorem
confirms the unbounded character of supersymmetric trajectories. The
evaluation of (\ref{eq:wirial_cl}) for a given trajectory shows that
the value of the time derivative of the virial explodes as the
particle escapes through the valley. Figure
\ref{wirial_klasyczny_susy} depicts theses results.
\par
    The findings obtained in this part show that the behavior
of classical trajectories and quantum states is analogous. For the
bosonic system quantum states are localized and classical
trajectories are bounded. On the contrary, for the supersymmetric
case, the majority of states are nonlocalized. The corresponding
classical motion is a trajectory which escapes through the valleys.
The localized supersymmetric quantum states are accompanied by
closed classical orbits. So, the differences observed between
quantum systems are also found on the classical level. This pattern
might indicate the existence of classical manifestation of
supersymmetry.

\subsection*{Conclusion} \label{sec:podsumowanie}
\par
    In this paper we investigated the properties of a system
proposed by de Wit, L\"uscher and Nicolai, in both quantum and
classical regimes. By comparing it with the bosonic system we
observed several interesting facts.
\par
    We started by introducing the quantum hamiltonians of the
bosonic and supersymmetric systems with the potential with four flat
valleys. Using the Born-Oppenheimer approximation we succeeded in
solving both systems. We concluded that a bosonic quantum particle
moving in one of the valleys is exposed to an effective potential
barrier which prevents it from escaping. Thus, all states are
localized. Contrary, the supersymmetric quantum particle can enter
at any depth into the valleys. The supersymmetric spectrum consists
of states from the continuous spectrum and, coexisting with them,
some localized states. Then, we introduced the hamiltonians of the
classical counterparts of these systems. For the supersymmetric
system we obtained the equations of motion which describe a particle
with spin which precess around some vector field. The interesting
feature is that this field is tangent to the contour lines of the
potential. Next, we described the cut-off method and some criteria
for testing the reliability of numerical results. The symmetries of
considered systems, discussed subsequently, fully explain all
degeneracies of exact energies. The comparison of the bosonic and
supersymmetric spectra confirmed the hypothesis based on the
approximated solutions on the character of quantum states. We
observed the noteworthy realization of the supersymmetric localized
states as deformations of the energy dependence on the cut-off. We
then turned our attention to the classical regime. We descibed the
behavior of trajectories of the classical bosonic and supersymmetric
systems. It turned out that the character of the trajectories
corresponds to the character of quantum states. Bosonic trajectories
are bounded, whereas the supersymmetric ones escape through the
valleys of the potential which can be due to the spin precession.
The above results show that the differences between quantum states
are also present on the classical level.
\par
    The supersymmetric potentials with flat directions appear also in some much more sophisticated
physical problems, i.e. in systems connected with Supersymmeric
Yang-Mills Quantum Mechanics (SYMQM). The investigated system is one
of the simplest models possessing many of the properties of SYMQM. A
deeper analysis can help to understand, for example, the coexistence
of the continuous and discrete spectrum in theses systems. Moreover,
the surprising effect of spin precession on the motion of a
classical, supersymmetric particle which enables it to escape
through the flat directions deserves a further attention.

\subsection*{Acknowledgments}
I would like to thank prof. J. Wosiek for invaluable help out with
this paper. This work is partially supported by the grant No. P03B
024 27 (2004-2007) of the Polish Ministry of Education and Science.

\end{document}